\documentclass[12pt,preprint]{aastex}
\begin{document}

\title{Challenges facing young astrophysicists}

\author{
N.L.Zakamska, A.E.Schulz, K.Heng, M.Juric, B.Kocsis, M.Kuhlen, R.Mandelbaum, \\
J.L.Mitchell, M.Pan, D.H.Rudd, G. van de Ven, Z.Zheng \\
{\sl School of Natural Sciences, Institute for Advanced Study, Einstein Dr., Princeton NJ 08540}
}

\begin{abstract}
In order to attract and retain excellent researchers and diverse individuals in astrophysics, we recommend action be taken in several key areas impacting young scientists:
\begin{itemize}
\item Maintain balance between large collaborations and individual
  projects through distribution of funding; encourage public releases
  of observational and simulation data for use by a broader community.
\item Improve the involvement of women, particularly at leading
  institutions.
\item Address the critical shortage of child care options and design reasonable profession-wide parental leave policies.
\item Streamline the job application and hiring process.
\end{itemize}
We summarize our reasons for bringing these areas to the attention of the committee, and we suggest several practical steps that can be taken to address them. 
\end{abstract}


\section{Introduction}
\label{sec_intro}

It is an exciting time to be a young researcher in astrophysics. The appeal of our field stems from its myriad attractions not always available in other branches of science. Astrophysics affords the opportunity to answer fundamental questions about the nature of the physical world. In contrast with more mature fields such as condensed matter or particle physics, astrophysics is compact enough that researchers can attain the frontier and conduct meaningful scientific inquiry without first investing the better part of a decade in the development of skills and background knowledge. One integral part of the astrophysics culture that draws bright ambitious students is specifically that this applied ``on the job'' training results in publications and substantial contributions to science.

Astrophysics also has a unique social structure that is conducive to attracting a more even demographic spread than other physical sciences. Our community welcomes the citizens of dozens of countries, and we are better gender-balanced than any other of the physical (hard) sciences. Partly this is due to the fact that the research of women and people of many ethnic backgrounds has appeared in the standard astronomical literature throughout the history of our subject, and partly it is due to concerted efforts that are made to draw women and minorities into the upper levels of academia. Furthermore, young astrophysicists in the United States -- much more so than in other countries -- are for the most part encouraged to seek out and lead their own topics, and it is common for junior faculty to be hired based on research programs they conceived and developed. At the same level of the academic ladder (e.g., at the postdoctoral stage), US astrophysicists enjoy more seniority, independence and recognition than their colleagues elsewhere.  

Nevertheless, there are several aspects of the astrophysics culture that threaten our ability to attract and retain an evenly sampled population of bright academics. The authors of this white paper, who are all postdoctoral fellows at the Institute for Advanced Study, have collectively identified obstacles that are likely to contribute to attrition of dynamic young researchers. The issues that we mention primarily affect graduate students, postdocs and junior faculty level academics. Among the twelve authors, six members have had a baby during their post-doctoral tenure and three had children previously during graduate school. Nine of the authors are married or engaged, and five are experiencing a significant ``two-body problem'' in the search for their next position. Six are not US citizens. Four of the authors are women. As a group, we have engaged in research that included acquisition and analysis of the data from radio to X-ray wavelengths, survey science and archive mining, numerical simulations on all scales, as well as analytical theory, and many of us switched between these topics more than once. 

We have identified four areas that are either in danger of
deteriorating and require vigilance on the part of the community to
sustain, or are in need of improvement. In section
\ref{sec_individual} we discuss maintaining astronomy as a field of
individual achievement.  In section \ref{sec_women} we address the
status of women in astrophysics. In section \ref{sec_children} we
address the issues of child care and parental leave. In section
\ref{sec_hiring}, we scrutinize the defects in the hiring process at the postdoctoral and junior faculty level.

\vspace{-0.1in}
\section{Astronomy as a field of individual achievement}
\label{sec_individual}

Progress in observational as well as theoretical astrophysics is frequently accomplished by single authors or small groups of authors, which has been noted by \citet{whit07} as a critical asset to retaining scientists who enjoy personal recognition for their achievements and require such recognition for career advancement. At the same time, collaborative efforts are becoming increasingly necessary to take advantage of large observational and computational resources, which are inherently inaccessible to the individual. Such collaborative projects have yielded important advances in astrophysics and are bound to be an important avenue of research in the next decade. 

The needs of the individual astronomer can be balanced with the field as a whole by carefully choosing the types of large projects which are performed and the way those projects are implemented. A greater priority should be given to multi-purpose or multi-focus projects, including surveys, where the quality of the data and its selection is not governed by any single research goal. The usefulness of such data is not limited by their original intent but by the creativity and ingenuity of the scientists who have access. Rigorous project management and publication policies\footnote{A successful example is the SDSS, http://www.sdss.org/policies/pub\_policy.html} are absolutely necessary to ensure a productive experience for young researchers within large collaborations, while at the same time giving credit to those who made the project possible in the first place. 

However, the use of any survey or collaboration dataset will remain limited if it is not available to the broader community. It is the public release of data that uniquely distinguishes observational astronomy from other experimental disciplines such as particle physics. While this critical step makes astronomical research reproducible -- true to the original meaning of the scientific enterprise -- it also allows individuals with limited resources to conduct valid and interesting observational work for the first time in history. Public releases of all observational data (after a reasonable proprietary period) acquired by collaborations or telescopes which receive financial support from the community should be the norm of the field. Furthermore, committees and funding agencies should recognize that additional resources are necessary to ensure prompt and high-quality data release and user support. Grants for public data releases can be made available for privately funded collaborations who would otherwise not have incentive to make such commitment. 

Computational astrophysics will increasingly face issues analogous to those of observational astronomy in the next 10 years, as {\it petascale} computing platforms become commonplace. While NASA Great Observatories have set firm standards for public availability of data, public releases of codes and simulation data are relatively new ground for astrophysics (although already several existing codes have become ``the industry standard'' because they were made publicly available and benefited from community feedback).  Unlike the case of observational data, there are no widely accepted standards for simulations or simulation outputs (although possible approaches have been discussed, e.g., by \citealt{shaw04}). The commitment needed to make these resources public is very substantial and includes both infrastructure to host codes and/or simulation products and manpower to maintain the products and interface with the community (\citealt{jans06} describes one such successful repository implemented in particle physics). Such costs need to be acknowledged in distribution of funds. Because there is rarely career-advancing incentive to publicly release simulation codes and data products, we propose creating separate grants to support public release. Release of simulation code and/or data should be regarded as a benefit to the broad astronomical community in much the same way as the release of observational data.

An attractive aspect of our culture is that astronomers frequently develop breadth as well as depth by delving into more than one sub-area. By maintaining this practice, the field appeals to those intellects that appreciate the interconnected nature of astrophysical problems. Ultimately this breeds a class of scientists with a highly developed intuition for the context surrounding their individual fields. Such role models will continue to lure the best and brightest candidates. {\bf The community should aim to achieve and maintain a balance between large collaborative endeavors and individual programs, and between investments into observational, computational and theoretical projects, thereby preserving the breadth of astronomical research.}

\vspace{-0.1in}
\section{Status of women in Astrophysics}
\label{sec_women}

Much has been said (but more remains to be done) about
under-representation of women in physical sciences in general and in
astronomy in particular\footnote{See, e.g., documentation from the AAS
  Committee on the Status of Women, http://www.aas.org/cswa/}. From
our perspective, this is indeed a problem, and one worth continuing
attention from the community, for the following reasons:
\begin{itemize}
\item First, there is now an understanding that women are
  under-represented not because of lacking intellectual capacity but
  because of social impediments. The additional hardship, whatever its
  source, is therefore ``unfair'' and should be eliminated.
\item Second, there is a concern that talented individuals are being driven away from the science disciplines, and therefore the field loses as a whole. 
\item Third, there is a general perception that gender diversity is important and makes for a more pleasant working environment. For example, lack of female role models is a significant deterrent for young female scientists entering the field, reinforcing the second issue listed above. 
\end{itemize}

Significant improvements in women's career advancement have been
achieved in the recent years, with astronomy clearly ahead of the
curve among other physical sciences. Indeed, representation of women
among faculty members is higher in astronomy than averaged over all of
physics (17\% in astronomy vs 13\% in physics as a
whole\footnote{Statistics provided by the American Institute of
  Physics,
  \\http://www.aip.org/statistics/trends/highlite/women3/figure1.htm,
  figure2.htm}) and has been on the rise in the recent years. The
difference is even more dramatic among the younger generation (28\% of
assistant professors in astronomy are women, compared with 17\% in
physics). 
It is clear that somehow astronomy departments have created
a significantly more women-friendly environment than departments in
other physical disciplines. One possibility is that astronomy is
benefiting from the historically higher representation of women, and
the availability of role models leads to higher retention rates for
young female scientists. On top of that, the astronomical community
benefits from being educated on the issue of gender biases (e.g.,
through the work done by the American Astronomical Society), and many
professionals have been involved in programs designed to improve
participation of women -- and these programs are working.

While these are encouraging trends, many problems remain. In particular, progress is significantly slower at top-ranked departments (among the top 20 physics departments, only 12\% of assistant faculty are women, compared with the 17\% for the field as a whole\footnote{http://www.aip.org/statistics/trends/reports/womenfaq.htm}). Very few full professors are female (6\% in physics and 11\% in astronomy). {\bf Forty-three percent of all physics departments have no women at all on their faculties}\footnote{http://www.aip.org/statistics/trends/highlite/women3/faculty.htm}, which is a very serious problem considering that many astronomers start by receiving their undergraduate education in physics. Furthermore, many astrophysicists are employed by physics departments not only in small colleges, but also in major PhD granting institutions. Therefore, the statistics for astronomy departments in the previous paragraph are misleading for astrophysics in general. Many factors that present obstacles for women's participation in sciences have been identified, including hidden gender biases in hiring and evaluation (\citealt{urry08} and references therein), as well as discouragement of female undergraduates linked to the aforementioned lack of female role models \citep{xie03}.

One major obstacle academics of both sexes frequently face is the tension between having a family and maintaining the competitive curriculum vitae necessary for attaining a desirable academic position. This issue has become particularly acute for women in astrophysics, as the standard duration of the postdoc position gradually creeps from three to six or seven years. The conventional wisdom has been to have children later in life after tenure has been achieved. While this may have been possible for previous generations, for most families this goal is now unrealistic.  However, having children during graduate school or a postdoc presents its own set of challenges and puts undue strain on families. Both the academic and biological clocks are ticking, and frequently the biological clock precipitates one or both partners leaving academia.

We would like to postulate, based on our own experiences and anecdotal evidence, that this is one of the most significant issues driving women from the field. For example, one of us, on her recent trip to a highly ranked astronomy department, had an informal meeting with female graduate students, who at that time were nearly 50\% of the graduate students in the program. Despite their (unusually) high representation in the graduate student population, every single woman at this meeting identified balancing family and career as a daunting problem potentially threatening her future participation in the field. Sociological research supports our assertion:
\begin{itemize}
\item \citet{maso02, maso04} study the effects of having children on science careers and the effects of having science careers on establishing a family, and conclude that there are significant gender differences. {\sl ``Women, it seems, cannot have it all while men can.''}
\item \citet{urry08}, while underscoring that the availability of child care is not the only factor affecting status of women in sciences, acknowledges that {\sl ``Childbirth has the effect of removing women from full-time work, to the long-term detriment of their careers. It is certainly true that there are too few high quality child care options available[...]''}
\item \citet{xie03} find that {\sl ``Short-term slowdowns [such as maternal leave] can have a very significant negative effect on a career overall.''}
\end{itemize}

We firmly believe that a significant improvement in women's participation in sciences can be reached if and only if {\bf there is a significant improvement in availability of child care and in policies and attitudes regarding parental leave}. While a few simple low-cost adjustments may provide some relief, a qualitative change in the situation will require financial investment. Some individuals within the astronomical community (sometimes with support of funding agencies) have been highly effective in public outreach programs encouraging participation of high-school and undergraduate female students, as well as in organizing meetings and publications to educate the members of the profession about the status of women. However, improvements in child care and parental leave policies cannot be easily achieved through individual contributions, but are well within reach of funding agencies -- as long as these issues are put on the high-priority list for the profession. With its relatively high participation of women and high interest in elimination of gender biases, the astronomical community is in an excellent position to provide a head start in this area, and such changes would benefit not only women in astronomy but men and women across all academic disciplines. The specific concerns regarding child care and parental leave are the subject of the next section. 

\vspace{-0.1in}
\section{Child care and parental leave}
\label{sec_children}

While it is not practical to expect the nature of the academic
hierarchy to change, there are some steps that could 
mitigate the strain felt by academics attempting to juggle careers and
children. We recognize that the lack of high-quality child care is a 
problem on a national level, which
does not pertain solely to astrophysics. The current reality is that
this issue is not on the federal political agenda, so it is not
tenable to expect federal funding or government programs to solve this
problem. Thus, the solutions we propose will not be free of cost
either to the individuals or the institutions, but we encourage the
community to seriously consider the advantages of making these types
of investments. Some institutions are already considering important
policies to make academic careers more family-friendly. The following
is based on proposed additions to the University of California's
family policies \citep{maso04} that we highly endorse. We hope that
collectively, institutions will make policies like these more the norm
than the exception.

\begin{enumerate}
\item A flexible part-time option for academics -- including tenured
  and tenure-track faculty, as well as postdocs and graduate
  students -- that can be used for limited periods as life-course needs
  arise.
\item A guarantee to make high-quality child-care slots available to
  academics.
\item An institutional commitment to assist new academic members with
  spousal-partner employment and other family-related relocation
  issues.
\item Re-entry postdoctoral fellowships that are designed to encourage
  PhDs who have taken time off from their careers to provide care to
  others to return to the academy.
\item Discounting of family-related resume gaps in the hiring of faculty.
\item Establishment of school-break child care and summer camps and of emergency backup child care programs.
\end{enumerate}

While we recognize that not every institution will be able to offer
these benefits, we expect institutions that do adopt such policies to
enjoy a competitive advantage\footnote{For example through the
  recently added ``benefits'' section in the job announcements on the
  widely-used AAS Job Register:
  http://members.aas.org/JobReg/JRIncludes/pubpol.cfm} in hiring and
retaining the best and brightest academics in the country,
particularly women faculty.

One of the most profound difficulties for academic families is the need to find good quality child care. Challenges faced by parents in academia include scanning the local area for appropriate facilities, adding themselves to waiting lists of often more than a year-long duration, being forced into contracts requiring more than eight hours of child care per day -- every day, and frequently commuting long distances because local child care options are full by the time parents receive an appointment at an institution.  We estimate that several weeks of valuable work time per child are lost in negotiating this clumsy process. A low-cost solution that every institution can implement on a short time scale is to maintain a registry of recommended child care providers and a program to help employees coordinate child care sharing. One essential ingredient of this solution would be to alert subscribers to the various registration deadlines and to provide updates on the remaining number of available slots. An intermediate-cost program that would assist young academic parents is a system of cheap on-site 24-hour ``emergency care'', which in practice would be limited to a few days per employee per year. More costly solutions include setting up an actual child care facility. As academics, we do enjoy certain flexibility in our profession, and therefore such a facility (or facilities recommended by the institution) should afford flexible part-time options, rather than a binary full-week-or-nothing policy. 

Several institutions that we know of have started implementing such
programs very recently (unfortunately, too recently for us to have
benefited from them), but the availability of these services remains
extremely scarce. The situation for graduate students and postdocs is
particularly dire because they are often not eligible for benefits
offered to faculty members and have much less financial leverage.
{\bf As more women -- and men with working partners -- enter the
  sciences, the availability of child care will continue lagging behind
  the demand unless special effort is made to address the issue.}

Another challenge facing academics with families is the ever-ticking academic clock.  Gaps in the curriculum vitae can make a candidate ineligible for junior faculty positions and certain fellowships. We propose that exceptions be made for individuals who have taken time off to raise children. The European Molecular Biology Organization offers special grants for young scientists who have taken time off to take care of the family\footnote{http://www.gunjansinha.com/nature\_medicine.htm}. American Physical Society offers a small year-long grant for women returning to active research after time off\footnote{http://www.aps.org/programs/women/scholarships/blewett/index.cfm}. The astrophysics community should seek funding to establish similar programs. 


Coming back into active research after time off is a very difficult enterprise, and therefore every effort should be made to help graduate students and postdocs maintain their careers while they are tending to their families. The leave, delay, and extension policies governing soft-money positions should be formalized. The rules should be established, followed, and widely circulated so that individuals are aware of options other than leaving the field. Graduate students are particularly vulnerable, because while some advisers are quite understanding of familial responsibilities, others may not be. Young researchers on soft-money positions should be granted unpaid leave or part time status on demand, while maintaining the available benefits. Extensions of these positions beyond the original contract date should also be considered when the individual chooses to start a family. This type of benefit is widely available in Europe, but is nearly non-existent in the United States. Most importantly, the community must be educated that young researchers should not be penalized for taking a reasonable time off when applying for their next position.

Tenure reviews can also be extremely harsh on academics raising young children. Many institutions now offer extended time for tenure review for both male and female parents, and we believe that this policy should become the norm of the profession.

Finally, financial contributions to child care and parental leave benefits should be added to the authorized expenses covered by infrastructure grants from funding agencies. Child care and parental leave are as valuable as computing and equipment costs in providing the necessary infrastructure to facilitate a researcher's ability to conduct research.

\vspace{-0.1in}
\section{Streamlining job applications and hiring process}
\label{sec_hiring}

Another topic that we would like to raise is the existing process of applying for a postdoctoral or junior faculty position in astronomy. The American Astronomical Society provides a remarkable service to astrophysics by maintaining the AAS Job Register\footnote{http://members.aas.org/JobReg/JobRegister.cfm} which is the primary -- and {\it de facto} the only -- means of advertising academic job opportunities in the field. In many other fields of science, such services are simply not available in any form and positions continue being advertised ``by word of mouth'', so astrophysics has an excellent head start in this area. It is also noteworthy that this resource is being increasingly used by foreign institutions.

Nevertheless, the process can stand to be further improved. For
example, at the postdoctoral application stage an inordinate amount of
time is taken by putting together a large number of applications with
the same content, but varying presentation requirements. A
straightforward solution is to standardize postdoctoral application
requirements. Furthermore, the transparency of the process would be significantly improved if the shortlist rankings and final decisions were announced in an organized and reliable way\footnote{The existing version of this process is the voluntary submission to the ``Astrophysics Job Rumour Mill'': \\http://cdm.berkeley.edu/doku.php?id=astrophysicsjobs\#astrophysics\_job\_rumour\_mill}. A more dramatic step would be to implement a third-party match-making solution. In this case, all institutions and all applicants submit their lists of preferences and the third party makes a final pairing\footnote{This problem has been shown to have a stable solution, e.g., http://en.wikipedia.org/wiki/Stable\_marriage\_problem}. Such algorithms have been designed for the medical field to match residents to hospitals, and are used nationally in the
US\footnote{http://www.nrmp.org/about\_nrmp/how.html}, in
Canada\footnote{http://www.carms.ca/index.html} and in the
UK\footnote{http://www.nes.scot.nhs.uk/sfas/About/default.asp} with
apparently satisfactory
results\footnote{http://www.dcs.gla.ac.uk/research/algorithms/stable/report.htm}.
Given the role of the AAS in the community, it could implement such a
scheme with relative ease.

The transition from the postdoctoral to junior faculty level is the least formalized, especially because there is no single duration of the postdoctoral stage applicable to all researchers. Those applicants who did not yet finish their postdoctoral position, but who would like to do so, often engage in a tedious negotiation with the employer institution about a possible deferral of a junior faculty position. Some institutions have been known to allow several-year deferrals, while others insist on an immediate starting date. Every effort should be made to accommodate such deferrals when they are requested, with the realization that the junior faculty member will greatly benefit from the additional training that they can receive free of job application pressure. However, in many cases there will be real financial reasons that would make such deferrals impossible. Therefore, at the very least the possibility of deferral and its duration should be part of the job description. (Of course, responsible behavior on the part of the applicant, such as honoring a binding contract, is necessary to make deferrals work.) The difference between timelines for appointments of junior faculty and postdocs presents an additional difficulty for those who apply for both types of positions. Those who accept a postdoctoral offer and are then offered a junior faculty position are often in an awkward situation, having to choose between deferring the faculty offer or rejecting the position they had already accepted. These issues may even put such applicants at a disadvantage in the considerations for faculty positions.  

Finally, one substantial difficulty with the final pairing of
applicants with positions are the needs of dual-career couples.
Couples attempting to obtain positions at the same or neighboring
institutions must prioritize differently when negotiating and
accepting offers.  It delays acceptance of an offer because an
applicant must wait until both people have heard from all the
institutions before the couple can jointly make a decision.  Any delay
in the application process negatively affects other shortlisted
individuals, who often accept other postdocs because they cannot wait
indefinitely for fear that they will not receive a job at all.
Furthermore, some applicants are secretive about their ``two-body''
problem, believing (often correctly) that they are less likely to
receive an offer if the institution perceives any conflict of
interest.  We recognize that there is no easy solution here
but note that coordinating the timeline for applications, 
interviews and offers will make it much
easier for such couples to make decisions and will streamline the
process considerably.  We also encourage institutions to develop
couple-friendly hiring practices.  Institutions in cities that host
many other institutions should consider forming joint programs to
coordinate job offers to couples who disclose the need to find
geographically compatible jobs.  In analogy to family-friendly
programs, we expect such hiring practices to present a very
considerable advantage for attracting talented couples.

\vspace{-0.1in}
\section{Summary}
\label{sec_summary}

In this contribution we have discussed several issues that affect many young scientists entering the field. 
\begin{itemize}
\item In the era of large surveys, expensive telescopes and super-computer simulations, astrophysics remains a field of individual achievement and broad expertise. These are wonderful features that need to be preserved through funding allocation and data archiving policies.
\item Participation of women in astrophysics and the experience of young scientists of both genders can be significantly improved if child care becomes easily available and if some reasonable parental leave policies are instituted.
\item Improvements can be made to streamline application process for junior positions, which would ease the burden for both applicants and institutions.  
\end{itemize}


\end{document}